\PassOptionsToPackage{unicode}{hyperref}
\PassOptionsToPackage{hyphens}{url}
\documentclass[
]{article}
\usepackage[a4paper, total={6in, 8in}]{geometry}
\usepackage{pdflscape}
\usepackage{tabularx}
\usepackage{fancyhdr}
\usepackage{tabu}
\usepackage[tableposition=below]{caption}
\usepackage{subcaption}
\usepackage{everypage}
\usepackage{adjustbox}

\usepackage{wrapfig}
\usepackage{blindtext}
\usepackage{multicol}
\usepackage{lmodern}
\usepackage{authblk}
\usepackage{amssymb,amsmath}
\usepackage{ifxetex,ifluatex}
\ifnum 0\ifxetex 1\fi\ifluatex 1\fi=0 % if pdftex
  \usepackage[T1]{fontenc}
  \usepackage[utf8]{inputenc}
  \usepackage{textcomp} % provide euro and other symbols
\else % if luatex or xetex
  \usepackage{unicode-math}
  \defaultfontfeatures{Scale=MatchLowercase}
  \defaultfontfeatures[\rmfamily]{Ligatures=TeX,Scale=1}
\fi
\IfFileExists{upquote.sty}{\usepackage{upquote}}{}
\IfFileExists{microtype.sty}{% use microtype if available
  \usepackage[]{microtype}
  \UseMicrotypeSet[protrusion]{basicmath} % disable protrusion for tt fonts
}{}
\makeatletter
\@ifundefined{KOMAClassName}{% if non-KOMA class
  \IfFileExists{parskip.sty}{%
    \usepackage{parskip}
  }{% else
    \setlength{\parindent}{0pt}
    \setlength{\parskip}{6pt plus 2pt minus 1pt}}
}{% if KOMA class
  \KOMAoptions{parskip=half}}
\makeatother
\usepackage{xcolor}
\IfFileExists{xurl.sty}{\usepackage{xurl}}{} % add URL line breaks if available
\IfFileExists{bookmark.sty}{\usepackage{bookmark}}{\usepackage{hyperref}}
\hypersetup{
  pdftitle={Academic Institutions in Multilateral Data Governance: Emerging Arrangements for Negotiating Risk, Value and Ethics in the Big Data Economy},
  hidelinks,
  pdfcreator={LaTeX via pandoc}}
\usepackage{longtable,booktabs,array}
\usepackage{calc} % for calculating minipage widths
\usepackage{etoolbox}
\makeatletter
\patchcmd\longtable{\par}{\if@noskipsec\mbox{}\fi\par}{}{}
\makeatother
\IfFileExists{footnotehyper.sty}{\usepackage{footnotehyper}}{\usepackage{footnote}}
\makesavenoteenv{longtable}
\usepackage{graphicx}
\makeatletter
\def\maxwidth{\ifdim\Gin@nat@width>\linewidth\linewidth\else\Gin@nat@width\fi}
\def\maxheight{\ifdim\Gin@nat@height>\textheight\textheight\else\Gin@nat@height\fi}
\makeatother
\setkeys{Gin}{width=\maxwidth,height=\maxheight,keepaspectratio}
\makeatletter
\def\fps@figure{htbp}
\makeatother
\ifLuaTeX
  \usepackage{selnolig}  % disable illegal ligatures
\fi

\newlength{\hfoot}
\newlength{\vfoot}
\AddEverypageHook{\ifdim\textwidth=\linewidth\relax
\else\setlength{\hfoot}{-\topmargin}%
\addtolength{\hfoot}{-\headheight}%
\addtolength{\hfoot}{-\headsep}%
\addtolength{\hfoot}{-.5\linewidth}%
\ifodd\value{page}\setlength{\vfoot}{\oddsidemargin}%
\else\setlength{\vfoot}{\evensidemargin}\fi%
\addtolength{\vfoot}{\textheight}%
\addtolength{\vfoot}{\footskip}%
\raisebox{\hfoot}[0pt][0pt]{\rlap{\hspace{\vfoot}\rotatebox[origin=cB]{90}{\thepage}}}\fi}

\providecommand{\keywords}[1]
{
  \small	
  \textbf{\textit{Keywords---}} #1
}

\title{Academic Institutions in Multilateral Data Governance: Emerging Arrangements for Negotiating Risk, Value and Ethics in the Big Data Economy}

\author[1]{Tsvetelina Hristova}
\author[2]{Liam Magee}
\author[3]{Emma Kearney}
\affil[1]{Institute for Culture and Society, Western Sydney University, Australia \authorcr t.hristova@westernsydney.edu.au}
\affil[2]{Institute for Culture and Society, Western Sydney University, Australia \authorcr l.magee@westernsydney.edu.au}
\affil[3]{emmak.etc@gmail.com}

\date{January 2023}

\begin{document}

\maketitle

\begin{abstract}
Data sharing partnerships are increasingly an imperative for research
institutions and, at the same time, a challenge for established models
of data governance and ethical research oversight. We analyse four cases
of data partnership involving academic institutions and examine the role
afforded to the research partner in negotiating the relationship between
risk, value, trust and ethics. Within this terrain, far from being a
restraint on financialisation, the instrumentation of ethics forms part
of the wider mobilisation of infrastructure for the realisation of
profit in the big data economy. Under what we term `combinatorial data
governance' academic structures for the management of research ethics
are instrumentalised as organisational functions that serve to mitigate
reputational damage and societal distrust. In the alternative model of
`experimental data governance' researchers propose frameworks and
instruments for the rethinking of data ethics and the risks associated
with it --- a model that is promising but limited in its practical
application.
\end{abstract}


\pagebreak

\begin{multicols*}{2}

\hypertarget{introduction}{%
\subsection{Introduction}\label{introduction}}

Data and big data, in particular, operates increasingly as a currency
integral to contemporary moral, knowledge and monetary economies. Its
production, collection and analysis are implicated in a multitude of
diverse practices, where data underpins epistemological and economic
practices and ideas that generate value which materialise along multiple
registers: as financial profit, epistemological insight, social benefit,
and institutional and political change. These notions of value can be
distributed unevenly across different social and economic domains ---
for instance, public and research institutions afford higher importance
to the epistemological and social gains, while corporate entities are
highly invested in the potential monetary returns from big data ---
there is increasingly an imperative for data sharing across both public
and private organisations. \emph{Big} data, ``generated continuously,
seeking to be exhaustive and fine-grained in scope, and flexible and
scalable in its production'' (Kitchin 2014a), is increasingly valued
less in its organisational containment---as privileged intellectual
property for example---and more through combinatorial possibilities of
sharing, linking and manipulation (e.g. Auer et al. 2007; Bizer, Heath
\& Berners-Lee 2009; Kitchin 2014b). Examples can range from highly
curated mechanisms for the sharing of personal and internet usage data
with advertisers (as is the case with big social media and platform
companies like Google and Meta), to data lakes necessitated by
public-private partnerships around projects for digitalisation and smart
urbanism (Scassa 2020), and further, to open data initiatives by
governmental institutions that provide varying degrees of access to
public datasets.

These arrangements for data sharing increase both the value and risks
associated with big data and, particularly, the risks of misuse and
privacy breaches that carry high-stake reputational hazards such as the
high-profile case of Cambridge Analytica. At the same time, the
mechanisms for addressing issues of data ethics and data governance in
the context of inter-organisational data sharing are still a challenge
for companies in their practical implementation and a relatively
understudied aspect in the social sciences, organisation studies and
information systems studies (Lis \& Otto 2020, 2021, Nokkala, Salmela \&
Toivonen 2019, De Prieëlle, De Reuver \& Rezaei 2020). Over the past
decade, concerns about data security and privacy have prompted action by
state and market holders of such sovereignty: legislative change by
national and transnational governments (e.g. European Parliament,
Council of the European Union 2016), and more stringent security
controls alongside apparently contrite apologies by technology
corporations like Google and Facebook in response to incidents of data
loss, hacking or exploitation (e.g. Zuckerberg 2019). As dramatic as
individual cases may be, their recurrence highlights the inadequacy and
immaturity of data governance regimes in an era of vast networked
information flows, across as well as within organisational borders.

Data sharing poses a challenge to the governance of data, which
traditionally links managerial and governance responsibilities to
structures and functions within the corporate organisation (Wende 2007).
Roles like data steward, data owner and data custodian are tightly
linked to how a particular organisation stores its data, what it
identifies as key data assets and how it aligns its data practices with
its particular domain of economic activity. Instead, data sharing across
organisational entities necessitates a different arrangement, which puts
the organisations participating in the data sharing partnership
\emph{themselves} in key roles like that of data stewardship, or
associated functions related to the ethical use or risk management of
data (Lis \& Otto 2021).

Our study contributes to this nascent field of research into the
governance of shared datasets between different entities, with a
particular focus on the role of research institutions in such
arrangements. We are interested in the functions adopted by universities
and research institutions in multilateral data partnerships in two
specific aspects: How are notions of value, risk and ethics negotiated
between these organisations and their partners within the data sharing
arrangement? And are research institutions afforded a particular role or
function?

Research institutions have long been guided by internal disciplinary
principles regarding their interaction with participants, the safety,
disclosure and availability of research data and the responsibilities of
individual academics and universities. With the growing amount of
digital data collected, analysed and stored by researchers, the
parameters of research ethics are increasingly shifting towards the
adoption of some form of data management or data governance principles
as part of the research process of conception, project description, and
ethics regulation. Some academic disciplines like biomedical research
have already pioneered research using big data and, subsequently, have
adopted principles for its management. Others, like the social sciences,
are still faced with the reality of negotiating two incongruent
frameworks of research ethics where principles of data management
enforced by IT departments and university risk management policies are
often seen as opposed to traditional discipline-imposed norms of
research ethics, as discussed in a 2018 forum in the journal Social
Anthropology (Pels et al. 2018). This tendency towards the incorporation
of data governance into the principles of scholarly research has been
accelerated by the growing imperative for open science collaborations
between academia and corporate partners, increasingly articulated
through the need of data sharing and collaborative data platforms
between universities and various production and service industries (see
Mirowski 2018, Fernández Pinto 2020).

Such developments have epistemic and political effects in academia,
where big data analytics feeds into fantasies of objectivity, control
and prediction (Halpern 2015, 2022), and generates new interdependencies
between risk, contingency and value. These interdependencies are
predicated on the expected gains unlocked by the adoption of practices
of datafication and data sharing (LaValle et al 2011). These gains are
part due to the nearly mystical role big data plays in harnessing and
channeling the destructive forces of risk and uncertainty (Parisi 2016).
At the same time, big data is not innocuous with respect to such forces
itself. Rather it exposes new forms of contingency that disturb regular
methods of risk management, and requires its models and comparatively
new practices of organisational governance (Tallon 2013).

\hypertarget{methods}{%
\subsection{\texorpdfstring{Methods }{Methods }}\label{methods}}

This article stems from a small internal and exploratory research grant
that looked at how different academic institutions navigate complex
arrangements of data sharing between public / non-profit and private /
commercial entities. Our research questioned how these data sharing
partnerships are organised to respond to the ethical and reputational
risks of data sharing, and what specifical roles the academic
institution adopts in these arrangements.

We have chosen four cases of existing research partnerships explicitly
formed around the imperative for data sharing, that have each produced
documentation explaining their approach to the ethics of data sharing
and the particular solutions adopted within their partnership. These
are: Social Science One, Insight Research Center, My Health Record, and
Big Data @ Heart. They operate at national, transnational or global
scale, involve both public and private institutions, and in varied ways
are at the forefront of emerging data governance and ethics practices.
All of the case studies are in Global North countries regions, a
particular methodological limitation of our study, and we hope that
future research on data governance partnerships will bring perspectives
from the Global South.

Each case is developed through publicly available documents, research
publications, media and websites. The detail of practices of ethics and
governance varies because of the different public interface each
partnership maintains via documents, news and forums, as well as how
those partnerships are represented in different media.

In our analysis of these partnerships we are interested in capturing key
components of an emerging political economy of data sharing, which we
believe to constitute important aspects of the economy of data
governance and data ethics. These are: who are the partners, what is the
organisational structure of the partnership, who funds the partnership,
what is the provenance of data curated within it, who can access it and
under what conditions. We use diagram and flowchart visualisations to
capture and represent the structure of each partnership, a
methodological decision that is part of the process of analysis but also
responds to calls for the use of visual tools for the analysis of data
partnerships as a way to illustrate the variations of emerging
inter-organisational forms (Lis \& Otto 2021).

\end{multicols*}

\pagebreak

% \begin{landscape}
% \pagestyle{empty}

\begin{longtabu}[]{@{} X X X @{}}
% \begin{longtabu}{| X | X | X |}
  \caption{Map of Data Governance Partnership roles and symbols} \\
\toprule
\endhead
\textbf{Role} & \textbf{Description} & \textbf{Symbol + Colour} \\
Data Partnership & The Data Partnership itself, which may comprise an
independent organisation (non-profit, etc), or a subordinate department
or institute with multiple accountabilities & Organisational Chart
Hierarchy \\
Data Manager & Responsible for hosting and managing data sets &
\adjustbox{width=5cm,height=4cm,valign=t}{\includegraphics[width=10cm]{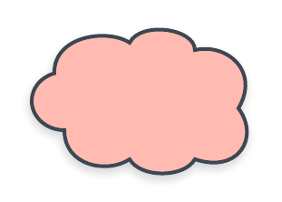}} \\
Funder or Investor & Funds the partnership - typically non-profit,
government or corporate actors &
\adjustbox{width=5cm,height=4cm,valign=t}{\includegraphics[width=10cm]{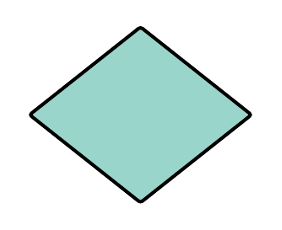}} \\
Data Governance Body (where independent of the partnership itself) &
Independent body responsible for aspects of governance, custodianship,
approving data requests and in some cases approving funding &
\adjustbox{width=5cm,height=4cm,valign=t}{\includegraphics[width=10cm]{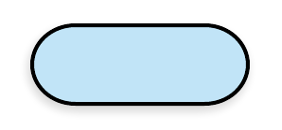}} \\
Data Subject & Citizens, patients and users who supply data to the
partnership, directly or indirectly &
\adjustbox{width=5cm,height=4cm,valign=t}{\includegraphics[width=10cm]{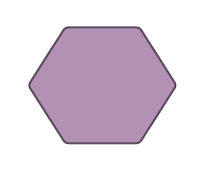}} \\
Data Client & Research institutions that request access to data and,
optionally, associated funding &
\adjustbox{width=5cm,height=4cm,valign=t}{\includegraphics[width=10cm]{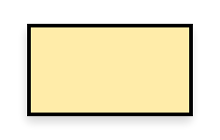}} \\
\bottomrule
\end{longtabu}

\thispagestyle{fancy}

% \end{landscape}
% \pagestyle{empty}

\begin{multicols*}{2}

\hypertarget{results}{%
\subsection{Results}\label{results}}

\hypertarget{social-science-one}{%
\subsubsection{Social Science One}\label{social-science-one}}

After the story of Cambridge Analytica broke in March 2018, leading to a
more than \$100 billion decline in Facebook's market capitalisation, the
company committed to changes in design and policy designed to exhibit a
commitment to privacy protection and user control over data (Tannam
2018). Social Science One (SS1), a not-for-profit Limited Liability
Company established by Harvard scholars for the purpose of managing
access of researchers to different Facebook datasets, is one of these
responses. Formed as an entity independent of Facebook, it emulates the
model of academic publishing in structure, with peer-review process and
editorial committee that acts as a gatekeeper in the interactions
between researchers and the social media platform, managing access to
Facebook data for research and control over the use of the datasets. SS1
is led by a commission chaired by its co-founders Gary King, founder of
the Institute for Quantitative Social Science (IQSS) at Harvard
University, and Nathaniel Persily, from Stanford Law School, and
consisting of sub-committees appointed by them.

SS1 relies on two key models of ensuring ethical compliance and
oversight: academic Institutional Review Boards (IRB) and a peer-review
process. All researchers applying to access data through Social Science
One must have approval via their University Ethics Review Board, and
their University must also take part in the data-sharing agreement to
ensure the institution is accountable for the individual researcher. The
importance of the safeguards afforded through institutional reputation
and IRB review processes is underscored by the financial support offered
to applicants from developing countries to undergo an ethics evaluation
process through independent IRBs. Proposals, along with the professional
ethical history of the researcher, are peer-reviewed, followed by a
further peer-review in the second stage of the grant process. Access to
Facebook data is tiered according to sensitivity levels and in the case
of high sensitivity data: ``researchers may need to develop analysis
code based on a synthetic data set and submit the code for automated (or
manual) execution, and where all data analysis (and literally every
keystroke) is subject to audit by us'' (Social Science One, 2018b).

\begin{figure*}
  \includegraphics{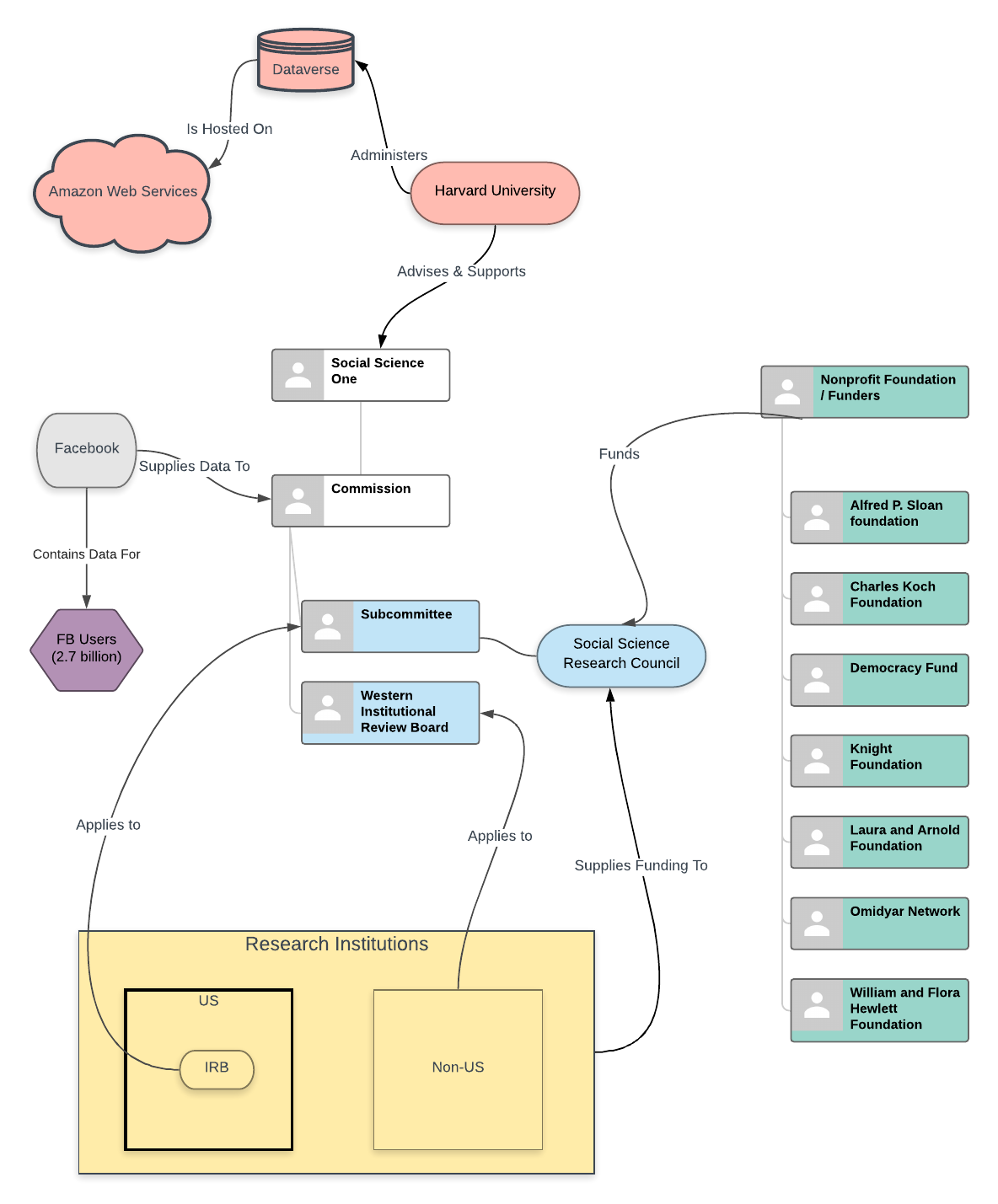}
  \caption{Social Science One data governance partnership}
  \label{fig:1}
\end{figure*}

Data released for use through Social Science One is de-identified and
stored in an open repository for research datasets,
\emph{Dataverse}\footnote{Data is stored @dataverse, which is a Harvard
  University open data repository open for worldwide use by researchers:
  https://dataverse.org/}, managed by the The Institute for Quantitative
Social Science at Harvard University. Differential privacy, a process of
adjusting data queries to obscure identifiable records, may be used to
protect the privacy of individuals while allowing researchers to access
data (Social Science One, 2018b). In addition to the governance model
for accessing funding and Facebook datasets, Social Science One offers
successful grant recipients a pre-peer review process for research
outputs that promises to fast track the time it takes for academic
papers to get published. Termed a ``peer pre-review'' process by
Harvard's Institute for Quantitative Social Science (IQSS), it
invites---and offers incentives to---senior scholars to scrutinise and
rate publications ahead of submission, ensuring the results comply with
a standard for replicability (King 1995).

In terms of governance structure, this partnership is centralised and
hierarchical (van den Broek and van Veenstra 2015), with Facebook
exercising exclusive control over the datasets that are released for
public research use. While the access to data is dominated by the
company, liability and accountability for research use is distributed
and outsourced to different institutions, such as IRBs and universities,
which put their reputation up as a guarantee in supporting proposals and
partnering with Social Science One. Risk is managed through tested
research principles and procedures---peer-reviews, replicability,
transparency, and independent research---a sign that the collaboration
leans heavily on the legitimacy of academic practices conducted by
partner universities. Reputational risk management has been organised
around existing models of risk and ethics in academic institutions, and
safeguards against unethical data use also draw upon the academic
institution for advice and oversight.

The funding for the Social Science One project comes from a range of
non-profit foundations: Laura and John Arnold Foundation, The Democracy
Fund, The William and Flora Hewlett Foundation, The John S. and James L.
Knight Foundation, The Charles Koch Foundation, Omidyar Network's Tech
and Society Solutions Lab, and The Alfred P. Sloan Foundation to the
Social Science Research Council. Distributed across philanthropic
sources, this funding model seeks to ensure academic independence from
Facebook, and alongside university endorsement, disperse risk and
accountability across multiple partners.

\begin{figure*}
  \includegraphics{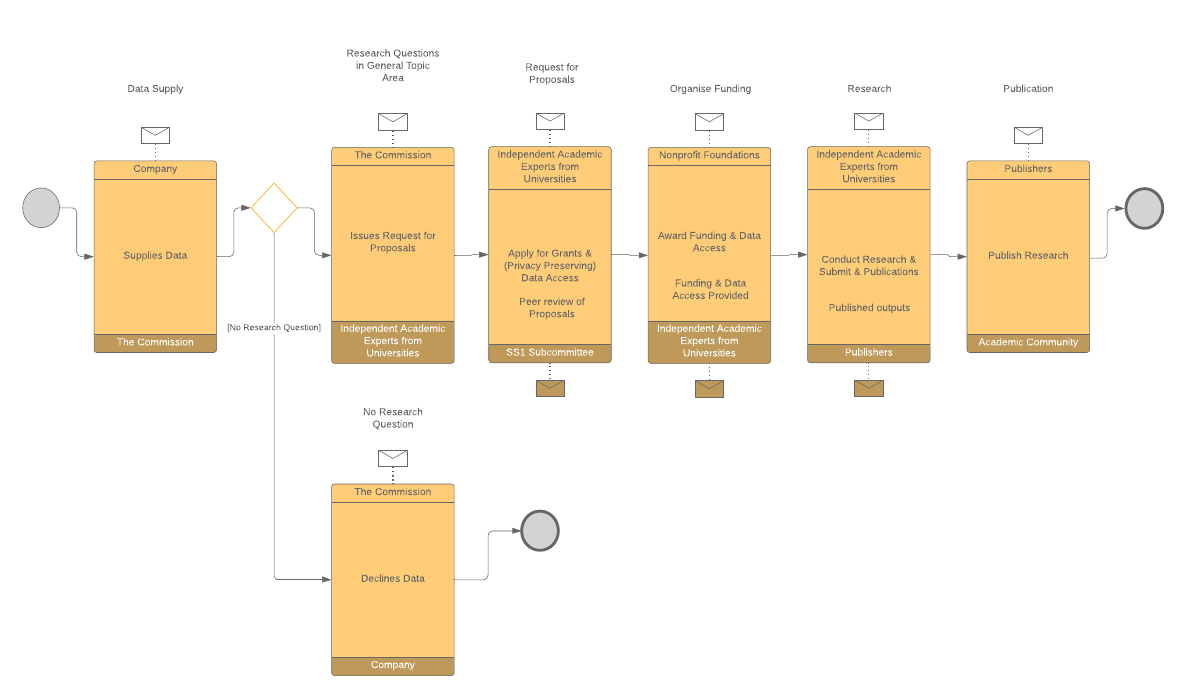}
  \caption{Social Science One data review process}
  \label{fig:2}
\end{figure*}

\hypertarget{my-health-record}{%
\subsubsection{\texorpdfstring{My Health Record
}{My Health Record }}\label{my-health-record}}

My Health Record (MHR) is an Australian Federal Government initiative to
centralise electronic health records in a standardised interoperable
database, containing the medical records and information of Australians
to enable data exchange and use between medical, research, and corporate
institutions. This database should be accessible by patients and medical
professionals providing treatment and consultation. Providers of digital
access services, such as appointment-booking apps, have conditional and
limited access to the database.

MHR is designed as a platform that allows uploading, storage, and
editing of health information on a system managed by the Australian
government. This presupposes a complex and flexible arrangement for
managing the rights to access the information of each patient.
Provisions for regulating access to the database from different entities
include the above-mentioned users: patients, healthcare providers,
mobile applications and researchers. MHR data access and use with these
different entities are regulated in different ways.

\emph{Primary (Clinical, Medical) Access and Use}

As shown in \emph{Figure 3}, the Australian government, represented by
the Australian Digital Health Agency, has sole authority over the My
Health Record platform and, to the extent to which access and use are
concerned, their conditions are set by the governmental institutions
through different legislative acts and regulations. This partnership is
thus dominated and regulated by the nation state, which attempts to
manage the access of private companies, healthcare providers, and
patients to MHR data.

\begin{figure*}
  \includegraphics{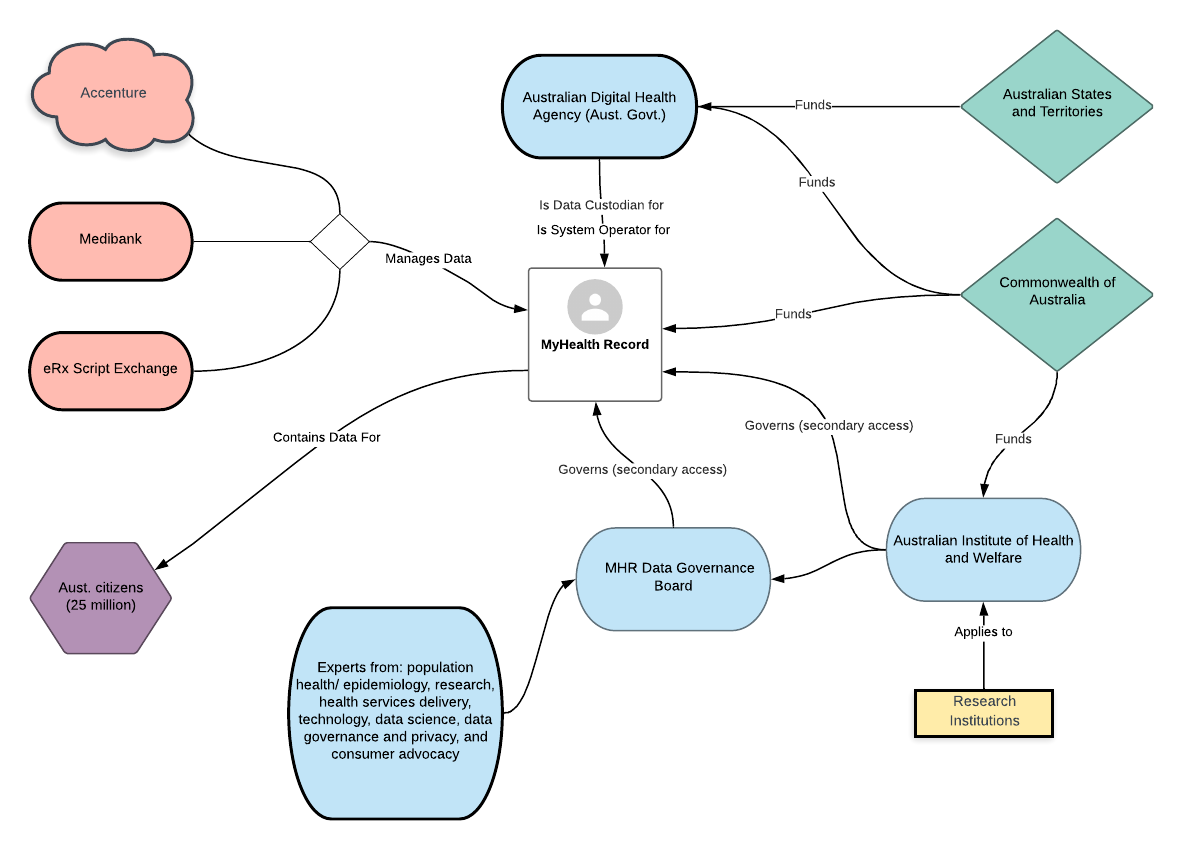}
  \caption{My Health Record data governance partnership}
  \label{fig:3}
\end{figure*}

My Health Record data is stored on registered repositories, which
include the Medicare\footnote{Medicare is the Australian Government
  funded Universal Health Care System.} repository for information
concerning medical conditions and treatment, and the eRx Script Exchange
repository for medical prescriptions (My Health Record 2018). Although
medical institutions can keep local copies of the documents related to a
patient, the My Health Record database is cloud-based. Cloud storage,
infrastructure, and maintenance are subcontracted to Accenture Australia
Holdings (Office of the Australian Information Commissioner 2014) and,
under the regulations for MHR, data is stored within the territory of
Australia (Commonwealth of Australia as represented by the Department of
Health 2018). The Portal Operator Registration Agreement, which
regulates relationships with companies that provide mobile and web
access to MHR data, also prohibits the operators to store or copy data
on their servers (Australian Digital Health Agency, n.d.). Despite this
provision and, due to the fact that Accenture is a multinational
enterprise, MHR has been exposed to criticism and scrutiny regarding its
security, which has been similarly vulnerable to breaches in other
countries and regions of operation of the company (Grubb 2017).

The MHR platform is an Australia-wide database, which means that it
navigates different state and Commonwealth (federal) legislations and
policies. Its organisational governance structure, therefore, is
designed to provide a supra-state level oversight through which
universal standards and regulations are administered, while
acknowledging the various regional actors, and their duties, rights, and
responsibilities. MHR distinguishes between the regulations for data
sharing within the healthcare system and data sharing for secondary use,
i.e. research. In its primary use, data from MHR is under the auspices
of the Australian Digital Health Agency, acting in its capacity as Data
Custodian, and data is shared exclusively for the purposes of healthcare
and diagnostics. This new institution of the Data Custodian has been
established by the Australian government especially for the purposes of
protecting the security and privacy of datasets that are in the
possession of state institutions. The Agency also acts in the capacity
of System Operator, controlling access to the database of MHR through
contracts, audits, and a registration process for healthcare providers.

Governance and funding of the MHR system and the institutions
responsible for its management are shared between the Commonwealth and
different states and territories, in accordance with the
Intergovernmental Agreement on National Digital Health (Council of
Australian Governments 2016). The MHR system is 100\% funded by the
Commonwealth, whereas the ADHA funding is shared among the Commonwealth
and the states and territories according to the AHMAC formula. Under
this formula the funding is shared 50:50 between the Commonwealth,
states and territories. Further division of costs among states and
territories is negotiated and decided by the Australian Health
Ministers' Advisory Council (AHMAC), an advisory group to the Council of
Australian Governments (Council of Australian Governments 2009).

\emph{Secondary (Research) Data Access and Use}

For the secondary use of MHR data---i.e. use of de-identified data for
research---the role of the Data Custodian is played by the Australian
Institute of Health and Welfare (AIHW). For the secondary use of MHR
data, the Data Custodian's institution is supplemented by a special MHR
Data Governance Board. The Board consists of representatives of AIHW
(which is also the accredited linkage authority) and independent experts
from population health/epidemiology research, health services delivery,
technology, data science, data governance and privacy, and consumer
advocacy (Commonwealth of Australia as represented by the Department of
Health 2018).

The Data Governance Board is responsible for assessing applications,
de-identifying data, implementing linkages, ensuring ethics conformance,
and reporting data breaches. In the evaluation procedure there is a
provision to examine the applications for secondary use of MHR data for
their public benefit. However, this value of the use of data is
established not by consumers/patients but by the institutional bodies
established for the purpose---i.e. by government bodies. Patients can
opt out from participating, and there is an as-yet-unrealised provision
of dynamic consent---applied individually, per case and per project.

\hypertarget{big-data-heart-case-study}{%
\subsubsection{Big Data @ Heart Case Study}\label{big-data-heart-case-study}}

Big Data@ Heart is a big data-driven translational healthcare research
platform, consisting of nineteen partner organisations representing
patient networks, learned societies, SMEs, pharmaceutical companies, and
universities across the EU. It aims to maximise the potential of big
data research in the medical and pharmaceutical fields, linking together
different health record databases from participating organisations and
standardising their format and method of analysis in order to yield more
precise and representative results and improve the efficacy of future
drug development. It brings together data of 25 million subjects across
Europe.

BigData@Heart is one of four disease-specific projects that sits under
the umbrella of the Innovative Medicines Initiative 2(IMI2) Programme,
Big Data for Better Health Outcomes, a research consortium jointly
funded by the European Horizon 2020 Research and Innovation Programme
and the European Federation of Pharmaceutical Industries and
Associations (EFPIA). The current consortium originates from the
European Technology Platform on Innovative Medicines (2005---2009), led
by the pharmaceutical industry who partnered with universities and other
stakeholders to develop a strategic agenda to boost drug research and
development in Europe. That work led in turn to the formation of a
public-private partnership called Innovative Medicines Initiative One
(IMI1), launched in 2008. The success of that initiative enabled a
second and current iteration, funded until 2024.

\begin{figure*}
  \includegraphics{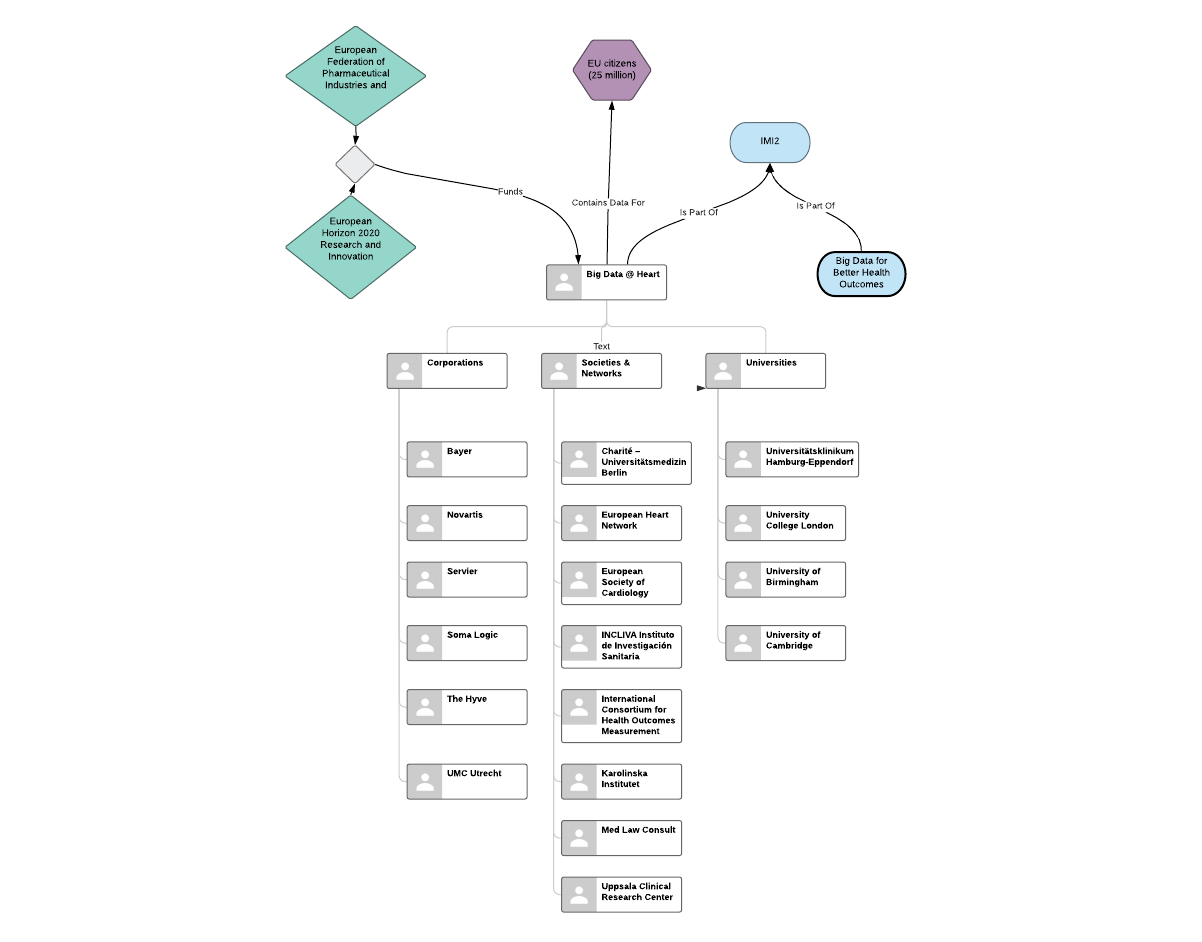}
  \caption{The Big Data@Heart data governance platform}
  \label{fig:big-data-heart}
\end{figure*}

Data is not stored by Big Data@Heart directly. Instead, it acts as a
platform hosting a metadata catalogue that researchers can use to
identify relevant datasets and contact relevant institutions and data
custodians to obtain access. These datasets allow the linkage of
Electronic Health Records (EHR) with omic, imaging, wearable and other
data.

IMI2, the umbrella organisation for both Big Data for Big Data@Heart and
Better Health Outcomes, acts as the governing body for research grants
for the consortium, with research topics decided through a governance
process that identifies high level priorities through the World Health
Organisation, European Union and the Pharmaceutical Industry. These are
then aligned to the IMI2 Strategic Research Agenda, done in consultation
with the IMI Scientific Committee and state representative groups. An
IMI2 annual work plan is developed based on agreed Annual research
priorities. Once these topics are identified, they go to the IMI
Governing Board for approval, after which calls for proposals are
distributed.

Big Data@Heart integrates data ethics and governance into its research
agenda. This is achieved through the design of the work process of the
consortium, which is divided into seven Work Packages. Work Package 7 is
concerned with `ethics legal and data privacy (governance, ethical and
legal aspects)' and is co-lead by academics from the University of
Utrecht and representatives of Bayer (University Medical Center Ultrecht
2017). The task of this group is to research and develop guidelines and
good practices for data governance and sharing across the partnership as
part of the ongoing project. The development of a governance framework
is seen as part of the knowledge production of the researchers involved
in Big Data @ Heart (van Thiel 2018), which shows how models of ethical
governance and trust in big data partnerships can participate in the
production of value in multiple ways. Seen not only as adding social
value to the legitimacy of big data analytics but as valued products in
their own right, such models are important outcomes of such
partnerships.

As part of the work under Work Package 7, Big Data @Heart conducted a
systematic review of literature and ethical guidelines to identify
principles and norms relating to data sharing in international health
research (Kalkman \emph{et al}. 2019). The authors identify four themes:
``societal benefits and value; distribution of risks, benefits and
burdens; respect for individuals and groups; and public trust and
engagement.'' As data is shared beyond the immediate clinical purposes
for which it was originally collected, issues of consent, third-party
use, security, privacy, and data sovereignty become important in the
work of Big Data@Heart, and these issues are further complicated by the
lack of harmonisation between different branches of local and national
legislation, on one hand, and between these forms of legislation and the
trans-European GDPR framework, on the other. IMI2 has conducted internal
social research into patient concerns on these topics, including opt-out
strategies, similar to those used in the Australian My Health Record
(MHR) case. The consortium also collaborates with different patient
organisations that provide advice on issues of privacy and social
benefit. Another aspect of the exploration of establishing ethics
guidelines has been discussion of a social license as a form of
contractual relation of trust between patients and the entities using
their data. Consent is operationalised here as a form of ongoing
engagement with how data is being used.

\hypertarget{insight-centre-for-data-analytics}{%
\subsubsection{Insight Centre for Data
Analytics}\label{insight-centre-for-data-analytics}}

The Insight Centre for Data Analytics is a research centre combining
expertise across four lead public universities: Dublin City University,
National University of Ireland, Galway, University College Cork, and
University College Dublin. As shown in \emph{Figure 5}, it also includes
four partner public institutions---Maynooth University, Royal Irish
Academy, Trinity College Dublin, and Tyndall National Institute---as
well as 80 industry partners. It is Ireland's national research centre
for data analytics, housing projects across four domains: Health and
Human Performance, Smart Communities and the Internet of Things;
Enterprises and Services; and Sustainability and Operations. The main
aim of the partnership is to produce high impact research in data
analytics, and to undertake projects that commercialise this research
and associated expertise for specific industry needs. It was launched in
2013, and succeeds and consolidates a previous research network of five
research centres (DERI, CLARITY, TRIL, Clique, and 4C). In 1998 the
Foundation launched a study of the Irish economy in order to identify
potential areas for future growth. Biotechnology and information and
communications technology were identified as two areas critical to the
future growth of the world economy, and the study concluded that ``world
class research capability in these two niches'' was a worthwhile
government investment to future-proof the national economy.

The Science Foundation for Ireland (SRI) provides Insight's core
funding, which continues until 2019, with a subsequent application for
funding has been submitted to SRI through until 2025. Insight
researchers are also lead investigators or co-investigators on a number
of EU-funded projects. The financial value of projects funded through
the centre are in excess of 100 million Euros.

\begin{figure*}
  \includegraphics{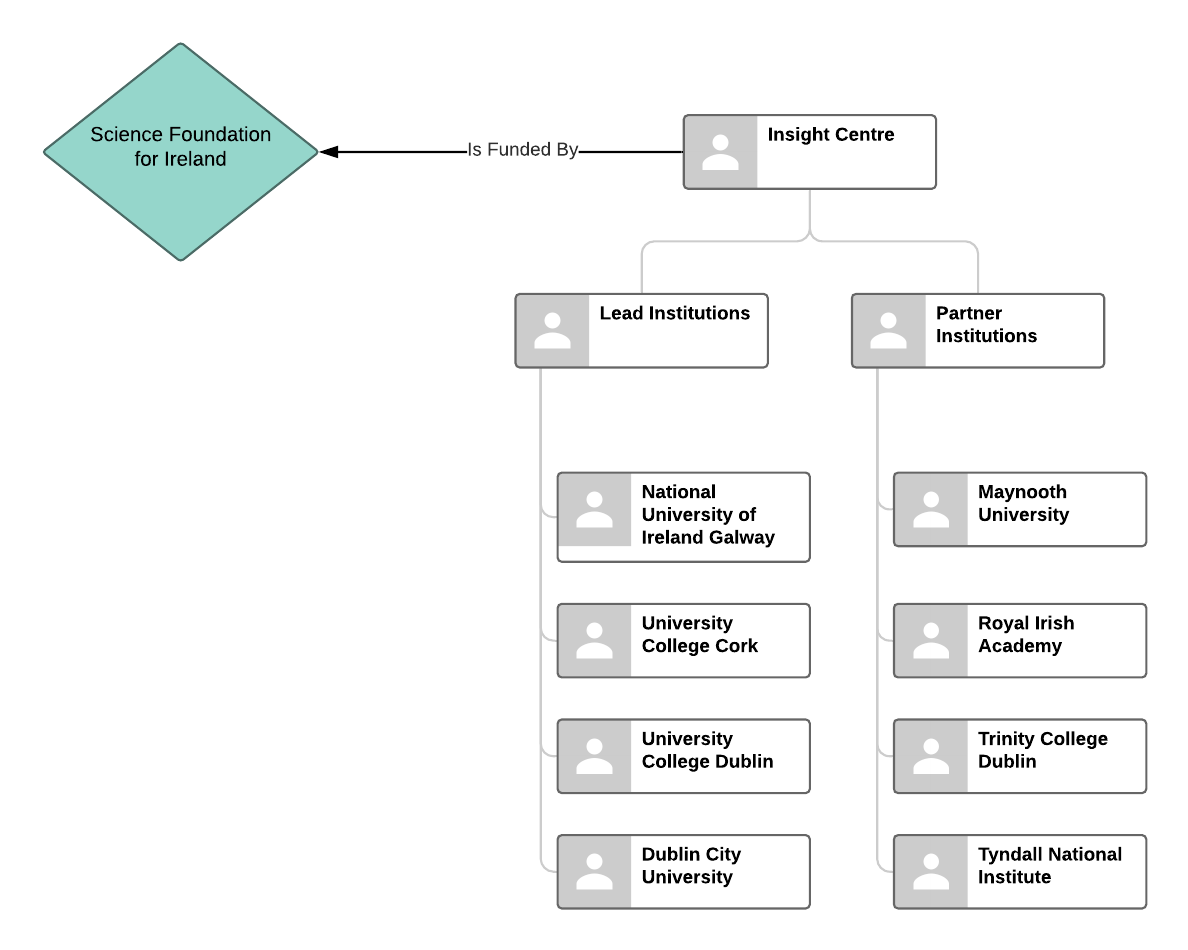}
  \caption{The Insight Centre for Data Analytics data governance partnership}
  \label{fig:insight-centre-for-data-analytics}
\end{figure*}

Insight is led by a team consisting of four researchers, and a CEO and
COO with industry management experience. The centre also houses a
scientific advisory committee and an industry advisory committee,
although their exact responsibilities are not outlined. Founded as part
of an Irish government strategy for innovation, the centre aims to
foster innovation and attract foreign investment in areas of data
analytics and the digital economy by providing a platform for connecting
academic and industry expertise. The centre serves as a research
platform, where public research institutions and industry collaborate in
multiple ways (Insight Centre for Data Analytics, n.d.).

Commercial spin-off companies resulting from projects conducted by
Centre researchers provide future employment for graduates in related
fields, and direct the commercialisation of research (Insight Centre for
Data Analytics, n.d.a). Aside from these commercial projects, Insight
conducts research that provide different options for collaboration
between the academy and technology industry, depending on types of
funding and concrete arrangements. Some of the projects are funded
through EU grants and, consequently, adhere to requirements and outputs
specific for EU-funded projects, such as open data access, while
retaining focus on research commercialisation.

The centre also provides possibilities for two types of industry
collaborations: those centred on a dedicated research question or area,
and others that respond to concrete commercial needs of a company
(Insight Centre for Data Analytics, n.d.c). While the first type of
partnership does not necessarily focus on commercial outcomes, the
second one is determined by the needs of the industry partner. Such
partnerships employ Insight scholars to analyse and make sense of
privately-owned customer or operational data. The conditions of use and
the protection of private data in these cases must comply with the
national legislation, while data collected by the researchers themselves
must also adhere to university ethics guidelines for research (Insight
Centre for Data Analytics, n.d.b). In return for sharing expertise and
facilities, researchers can access large multi-faceted collections of
social or other data that is beyond the means---and perhaps the ethical
remit---of universities themselves to acquire (Insight Centre for Data
Analytics, n.d.d).

Such arrangements offer a competitive research setting for developing
data science and analytics expertise. As there are multiple types of
funding and collaboration arrangements, no single model for data
governance exists across these different partnerships. However, the
description of Insight's four-year research collaboration with SAP gives
some idea about the possible arrangements. Over the course of this
partnership, SAP has funded PhD positions for researchers directly
involved in the research partnership, provided expertise and placement
for mentoring of the PhD students, and collaborated on preparation of
academic publications. The company also gave access to its SAP HANA
database for research projects (Faller 2018).

As a data sharing collaboration founded directly to foster an innovative
data economy in Ireland, Insight provides substantial benefits for both
the research institutions and their industry partners while avoiding
some of the obstacles of risk management across both academic and
industry settings. Researchers gain access to datasets and technology
that would otherwise be safeguarded through intellectual property and
privacy arrangements, and would be inaccessible for research uses.
Companies meanwhile can take advantage of reputational benefits, diverse
expertise and dedicated research teams for specific projects without the
costs and burdens of recruitment and project management.

One of the flagship projects of the centre---\emph{Magna Carta for
Data}---focuses directly on questions of risk, value, and ethics of big
data research in the data economy. A continually revised document that
ventures rather than proscribes principles for big data collaborations,
in its current form it suggests directions for debate and potential
changes in data sharing practice and legislation. Repeating the
often-referenced claim that big data is ``the new currency'' (Insight
Centre for Data Analytics, 2015, p. 6), Magna Carta engages directly
with many widely debated ethical topics: fairness, ownership, privacy,
and individual rights. To foster debate and develop the prospective bill
of data rights further, Insight has run workshops and hosted an online
platform where various data stakeholders can contribute and discuss
ideas (Insight Centre for Data Analytics, 2017).

In key respects, Magna Carta for Data makes explicit claims that must be
inferred from the other cases discussed here. Situating ethical concerns
in adjacent rather than oppositional relation to commercial interests,
it performs an instrumentalisation of ethics as simultaneously `lock'
and `key': a protection for data subjects that, once enacted, can also
look to unleash potential for further value generation, expanding fields
of knowledge and technology innovation. Ethics is established as a key
multifunctional instrument to secure trust from the ``originators of
data'' that will make possible future growth of the data economy
(Insight Centre for Data Analytics, 2015, p. 6).\footnote{The authors
  expand on the role of ethics in an economy of data exchange: ``Data is
  a valuable currency in research, industry and society. It potentially
  holds the key to better, safer, more efficient societal systems and
  much more besides. However a balance is needed. The temptation is
  ever-present for industry and others to exploit the data of
  individuals for profit and yet, without the \emph{trust and
  cooperation} of the originators of that data, any sustainable system
  of research would be impossible.'' (Insight Centre for Data Analytics,
  2015, p. 7 -- our emphasis).} Noting that fears of privacy violation
can act to hinder the benefits and innovation potential of big data
science and industry, the provisional bill of data rights also critiques
an exclusive and limited focus on privacy, which distract ``from the
wider issues around data ethics'' (Insight Centre for Data Analytics,
2015, p. 8). Such wider issues demand instead multilateralism, and
partnerships supply the ideal model for adjudicating on them: ``we can
balance the needs and rights of individuals in the Big Data age, but
collaboration and cooperation will be key'' (Insight Centre for Data
Analytics, 2015, p. 8).

\hypertarget{discussion-emerging-models-of-governance-in-big-data-research-partnerships}{%
\subsection{Discussion: Emerging Models of Governance in Big Data
Research
Partnerships}\label{discussion-emerging-models-of-governance-in-big-data-research-partnerships}}

With regard to the four case studies, our primary question is how issues
of data ethics and governance are operationalised and integrated within
a model of value extraction and within a specific organisational
structure. Summarised by key characteristics in \emph{Appendix 1}, the
four case studies here present collaborations between different types of
organisations---public, government, academic, and commercial---and this
means that notions of risk, value, and ethics are negotiated and made
interoperable across different organisational contexts and through
specific organisational arrangements. Key for our identification of a
new organisational form of a data sharing partnership is the
establishment of common data governance structures and procedures among
the partnering entities. The analysis of the case studies shows that
there are two emerging tendencies in the development of new
organisational forms of multilateral data governance, which we term
\emph{combinatorial} and \emph{experimental}.

The combinatorial approach to data partnership relies on utilising
existing organisational structures across industry, academia, and public
institutions, which are already associated with ethics oversight and
risk management. This approach can serve to distribute responsibilities
across domains but, conversely, could reinforce and solidify more
traditional divisions and relationships between ethics, value, risk, and
the organisational fields that operationalise them. One such possibility
is the delegation of tasks associated with the management of ethics and
related risk to academic or public institutions, which would then serve
as a `trust cushion' for commercial partners. The other tendency, which
we refer to as an experimental approach to data partnerships, concerns
the rethinking of the core relation between ethics, risk, trust, and
value within the partnership, which might or might not be directly
translated into a specific organisational form. Foregrounded explicitly
in some of the partnerships, this approach treats established
relationships between risk, value and ethics---evident in existing
legislation, research protocols, and institutional practices---as
fertile ground for re-examination. These two tendencies develop on a
continuum and, rather than being mutually exclusive, refer to different
levels of the organisation of data sharing, ethics, and value
extraction.

While both tendencies point to emergent innovative organisational forms
and the integration of different entities around the governance,
exchange, and valorisation of big data, the combinatorial characteristic
of data partnerships solidifies dependencies between function and form
in the operationalisation of ethics. This tendency is exemplified in the
delegation of the functions of data custodianship, ethics oversight, and
processual compliance to academic, public, and non-profit institutions
traditionally associated with transparency, ethics, and social benefit.
It necessitates a specific division of data management labour that
strips down functions and streamlines collaboration, and therefore
implies a degree of conservatism and formalism in addressing the issue
of ethics and associated risks. This combinatorial approach is less
concerned with ambitions to rethink and remake ethics in the age of big
data, and rather seeks to identify the proper entities to serve the
function of ethical oversight and compliance.

While each of the case studies exhibits a degree of combinatorial
organisational structure, insofar as they make use of already existing
organisational forms and institutions, Social Science One exemplifies
this combinatorial characteristic with respect to the ethics and risks
of big data sharing. Acutely defined by a background of trust breaches
and the reputational and economic risks associated with them, the
initiative is structured as a simultaneously centralised and
compartmentalised arrangement between the participating entities. While
data is exclusively managed by the social media corporation that
prepares datasets for release, the tasks of ethical oversight and
financially independent funding are delegated to entities that are
traditionally associated with such tasks. Ethical oversight is
guaranteed through the inclusion of Institutional Review Boards in the
preparation of the research proposals---a bureaucratic form that itself
has been criticised for being unable to respond to the challenges of big
data research (Leetaru 2016)---and later, during research dissemination
and assessment, by academic research and publishing institutions, while
the funding of research projects is steered through independent
non-profit foundations. Social Science One effects a complex design of
risk delegation and management, where Facebook separates into discrete
functions the realisation of value and the evaluation of ethics, and
outsources this evaluation in return for modest grant schemes and highly
restricted data access. This design points to a central characteristic
of the combinatorial data partnership: it is defined by a strategy of
operationalising existing organisational structures and optimising the
equations of value, risk, and ethics according to the differing
resources, aptitudes and purposes of those structures. That the
university ethics board becomes, in the case of Social Science One, the
receptor of functions related to ethics and reputational risk in these
combinatorial arrangements speaks to a conventional and conservative
institutional demarcation between ethical oversight and economic value
production.

The second combinatorial partnership example, MHR, configures the
``risk-value-ethics'' equation differently, and also integrates
suggestive experimental possibilities. A nationwide organisation and
centralisation of a vast dataset of personal data, it seeks to foster
economic and innovation benefits while regulating the relations, rights,
and duties of different actors in the digital healthcare industry. The
formation of MHR has taken place amid discussions about data storage,
sharing, and valuation between the Australian Productivity Commission,
which issued a report on data availability and use in March 2017
(Australian Government Productivity Commission, 2017), and other areas
of the federal government (Department of the Prime Minister and Cabinet,
2018). The Productivity Commission's report raises questions about
public benefit and economic gain---or in the terms of our analysis, risk
and value---associated with data sharing across different sectors and
institutions. It suggests an alternative and devolved approach to data
sharing that focuses on the possibility of individuals benefiting from
direct control over their data, rather than the traditional exclusive
concern with data privacy and security. This aspect of the report's
recommendations has been partially adopted by the federal government,
which has since decided to introduce consumer data rights as a
legislative concept and practice, with effect across specific schemes
such as the MHR (Australian Competition \& Consumer Commission, 2020).
In legislative terms, consumer data rights appear to be widening from a
purely protective framing---the right to privacy for example---to one
that also encompasses positive rights to control access through a series
of rules about consent granting, duration and withdrawals,
authorisation, and data holder accreditation (Australian Competition \&
Consumer Commission, 2019).

The Insight initiative proposes a more radical rethinking of risk and
ethics, proposing that institutions seeking to extract value from big
data would explicitly signal compliance with a charter or bill of
rights. By managing a platform and encouraging dialogue about the Magna
Carta for Data, Insight opens a field for experimentation and dialogue
that makes explicit aspect of the very calculus of ethics, risk and
value we have argued big data partnerships must negotiate. The
experimentalism of both Big Data@Heart and the Insight Centre should not
be overstated, but so far seem to exhibit a more nuanced approach to
such calculations, acknowledging the multiple registers of value and
benefit that may be realised through financial compensation, improved
health treatments, political representation and other advances in
research. `Value' is developed here implicitly as a quality that accrues
to individuals as much as institutions, and protective
aspects---relating to security vulnerability and privacy for
example---must be balanced against that capacity for accrual within
these calculations of risk.

\hypertarget{conclusions-reconfiguring-risk-value-and-ethics}{%
\subsection{Conclusions: Reconfiguring Risk, Value and
Ethics?}\label{conclusions-reconfiguring-risk-value-and-ethics}}

Our paper makes three contributions. First, in the four case studies
illustrated here data governance partnerships are identified as an
emerging organisational form. These develop in an environment that
technically, makes possible data sharing at scale, and organisationally,
recognises the value in doing so. Second, intrinsic to all such
partnerships is a specific relationship between risk, ethics, and value,
and we have sought to explore how, in the establishment of arrangements
for data governance, partnership actors strategise about how this
relationship is configured. Third, this relationship is not static
across partnership instances, and the partnership itself can be seen as
a device for rethinking and reconfiguring under limited experimental
conditions how it may be articulated on a continuum of combinatorial and
experimental characteristics.

As organisational form, the data partnership occupies a critical vantage
point in the unfolding terrain of the political economy of big data.
That terrain has geopolitical features we have only begun to gesture
towards here. In the EU, regulatory frameworks such as the GDPR have
begun to enshrine principles of accountability and compliance as a
necessary part of data governance for all interested parties, and the
implications of such transnational regulations inform the context and
discussions of data ethics in all four partnerships. Such frameworks
also intersect with national legislative frameworks, as well as
established approaches in ethics of data management and research in the
fields across which the four partnerships spread. Our overview of these
partnerships shows that legislative instruments form part of the
landscape within which value, risk, and ethics are negotiated, but they
are not the only means of defining and enacting data ethics and
governance.

What prospects exist for these arrangements? As other authors have noted
(such as Lis \& Otto 2020, 2021), the governance of shared datasets
requires emerging organisational solutions that are often established ad
hoc and can differ significantly from case to case. However, these can
gravitate towards replicating existing models of sector-specific
configurations for the governance of ethics, value and risk, as
exemplified by what we term combinatorial data governance arrangement,
or by partnerships adopting the extractive model of digital platforms
(Fernández Pinto 2020, Marelli, Testa \& Van Hoyweghen 2021, Srnicek
2017). The cases analysed here offer two possible avenues for academic
institutions partnering in such arrangements---either as a dedicated
guarantor of ethical oversight and the management of reputational risks,
or as a leader in the discursive and theoretical re-examination of the
relationship between risk, value and ethics in the sharing of large
datasets. It is, however, still questionable whether such scholarly
intervention have the potential to introduce new, more democratic, and
more innovative models of collaborative data governance in the face of a
growing pressure from public institutions and private industries to
position researchers as mere providers of research services and
guarantors of public trust.

\hypertarget{references}{%
\subsection{\texorpdfstring{\textbf{References}\\
}{References }}\label{references}}

Auer, S., Bizer, C., Kobilarov, G., Lehmann, J., Cyganiak, R., \& Ives,
Z. (2007). \emph{Dbpedia: A nucleus for a web of open data}. In
\emph{The Semantic Web} (pp. 722-735). Springer, Berlin, Heidelberg.

Australian Government Productivity Commission. \emph{Data Availability
and Use: Productivity Commission Inquiry Report}. (2017).
\url{https://www.pc.gov.au/inquiries/completed/data-access/report/data-access.pdf.\%20Accessed\%2010\%20February\%202020}{https://www.pc.gov.au/inquiries/completed/data-access/report/data-access.pdf.
Accessed 10 February 2020}.

Australian Competition \& Consumer Commission. Competition and Consumer
(Consumer Data) Rules. (2019).
\url{https://www.accc.gov.au/system/files/Exposure\%20draft\%20CDR\%20rules\%2029\%20March\%202019.pdf}.
Accessed 10 February 2020.

Australian Competition \& Consumer Commission. Consumer data right
(CDR). (2020).
\url{https://www.accc.gov.au/focus-areas/consumer-data-right-cdr-0}.
Accessed 10 February 2020.

Australian Digital Health Agency. Portal Operator Registration
Agreement. (n.d.).
\url{https://www.digitalhealth.gov.au/about-the-agency/freedom-of-information-foi/agency-plan/Portal-Operator-Agreement-_ViewOnlyAccess.pdf}.
Accessed 10 February 2020.

Bizer, C., Heath, T., \& Berners-Lee, T. (2009). Linked data: The story
so far. In \emph{Semantic services, interoperability and web
applications: emerging concepts} (pp. 205-227). IGI Global.

Commonwealth of Australia as represented by the Department of Health.
Framework to guide the secondary use of My Health Record system data.
(2018).
\url{http://www.health.gov.au/internet/main/publishing.nsf/content/F98C37D22E65A79BCA2582820006F1CF/$File/MHR_2nd_Use_Framework_2018_ACC_AW3.pdf}.
Accessed 10 February 2020.

Cooper, C. (2015). Accounting for the fictitious: a Marxist contribution
to understanding accounting's roles in the financial crisis.
\emph{Critical Perspectives on Accounting}, 30, 63-82.

Council of Australian Governments. Intergovernmental Agreement on
National Digital Health. (2016).
\url{https://www.coag.gov.au/sites/default/files/agreements/digital-health-iga-signed.pdf}.
Accessed 10 February 2020.

Council of Australian Governments. National Partnership Agreement On
E-Health. (2009).
\url{https://www.coag.gov.au/sites/default/files/agreements/digital-health-iga-signed.pdf}.
Accessed 10 February 2020.

Department of the Prime Minister and Cabinet, Australian Government. The
Australian Government's response to the Productivity Commission Data
Availability and Use Inquiry. (2018).
\url{https://www.pmc.gov.au/public-data/data-integration-partnership-australia.\%20Accessed\%2010\%20February\%202020}{https://www.pmc.gov.au/public-data/data-integration-partnership-australia.
Accessed 10 February 2020}.

De Prieëlle, F., De Reuver, M. and Rezaei, J., (2020). The role of
ecosystem data governance in adoption of data platforms by
Internet-of-Things data providers: Case of Dutch horticulture
industry.~\emph{IEEE Transactions on Engineering Management}.

European Parliament, Council of the European Union. Regulation (EU)
2016/679 of the European Parliament and of the Council of 27 April 2016
on the protection of natural persons with regard to the processing of
personal data and on the free movement of such data, and repealing
Directive 95/46/EC (General Data Protection Regulation). (2016).
\url{https://eur-lex.europa.eu/legal-content/EN/ALL/?uri=CELEX:32016R0679}.
Accessed 10 February 2020.

Faller, G. Insight and SAP's four year collaboration a huge success.
(2018).
\url{https://www.insight-centre.org/News/insight-and-saps-four-year-collaboration-huge-success.\%20Accessed\%2010\%20February\%202020}{https://www.insight-centre.org/News/insight-and-saps-four-year-collaboration-huge-success.
Accessed 10 February 2020}.

Fernández Pinto, M., (2020). Open Science for private interests? How the
logic of open science contributes to the commercialization of
research.~\emph{Frontiers in Research Metrics and Analytics},~\emph{5},
article 588331, p. 1-10.

Halpern, O., (2015).~\emph{Beautiful data: A history of vision and
reason since 1945}. Duke University Press.

Halpern, O., (2022). The future will not be calculated: Neural nets,
neoliberalism, and reactionary politics.~\emph{Critical
Inquiry},~\emph{48}(2), pp.334-359.

Kalkman, S., Mostert, M., Gerlinger, C., van Delden, J. J. M., \& van
Thiel, G. J. M. W. (2019). \emph{Responsible Data Sharing in
International Health Research: a Systematic Review of Principles and
Norms}. BMC Medical Ethics, 20(1), 21.

King, G (1995). Replication, Replication. \emph{PS: Political Science
and Politics}, 28, 444-452.

King, G. (2007). An Introduction to the dataverse network as an
infrastructure for data sharing. \emph{Sociological Methods \&
Research}, 36(2), 173-199.

King, G., \& Persily, N. (2018). A New Model for Industry--Academic
Partnerships. \emph{PS: Political Science \& Politics}, 1-7.

Kitchin, R. (2014). Big Data, New Epistemologies and Paradigm Shifts.
\emph{Big Data \& Society}, 1(1).

Kitchin, R. (2014). \emph{The Data Revolution: Big Data, Open Data, Data
Infrastructures and their Consequences}. Sage.

Innovative Medicines Initiative (IMI). \emph{The Innovative Medicines
Initiative and patients---a partnership.} (2016).
\url{https://www.imi.europa.eu/sites/default/files/uploads/documents/reference-documents/PatientBrochure2016.pdf}.
Accessed 10 February 2020.

Innovative Medicines Initiative (IMI). \emph{Big data for better outcomes (BD4BO).} A
Practical Toolkit for the identification, selection and measurement of
outcomes including in real world settings. (2018).
\url{http://bd4bo.eu/index.php/toolkit}. Accessed 10
February 2020.

Insight Centre for Data Analytics. \emph{Towards a Magna Carta for
Data}. (2015).
\url{https://www.insight-centre.org/sites/default/files/basic_pages_file/towards_a_magna_carta_for_data.pdf}.
Accessed 10 February 2020.

Insight Centre for Data Analytics. \emph{Magna Carta for Data---About}.
(2017). \url{http://magnacartafordata.org/about/}. Accessed 10 February
2020.

Insight Centre for Data Analytics. \emph{Commercialisation}. (n.d.).
\url{https://www.insight-centre.org/content/commercialisation}. Accessed
10 February 2020.

Insight Centre for Data Analytics. \emph{Data Protection Notice}.
(n.d.).
\url{https://www.insight-centre.org/Basic-page/data-protection-notice}.
Accessed 10 February 2020.

Insight Centre for Data Analytics. \emph{Insight for Business}. (n.d.).
\url{https://www.insight-centre.org/content/insight-business}. Accessed
10 February 2020.

Insight Centre for Data Analytics. \emph{Strategic Plan 2017---2025}.
(n.d.).
\url{https://www.insight-centre.org/sites/default/files/basic_pages_file/strategy_doc.pdf}.
Accessed 10 February 2020.

Lawler, M., Morris, A. D., Sullivan, R., Birney, E., Middleton, A.,
Makaroff, L., Knoppers, B. M., Horgan, D. \& Eggermont, A. (2018). A
roadmap for restoring trust in Big Data. \emph{The Lancet Oncology},
19(8), 1014-1015.

LaValle, S., Lesser, E., Shockley, R., Hopkins, M. S., \& Kruschwitz, N.
(2011). Big data, analytics and the path from insights to value.
\emph{MIT Sloan Management Review}, 52(2), 21.

Leetaru, K., (2016). Are research ethics obsolete in the era of Big
Data. \emph{Forbes.} (2015).
\url{http://www.forbes.com/sites/kalevleetaru/2016/06/17/are-research-ethics-obsolete-in-the-era-of-big-data}.
Accessed 10 February 2020.

Lis, D. and Otto, B., (2020). Data Governance in Data
Ecosystems---Insights from Organizations.~\emph{AMCIS 2020 Proceedings}.
12.\\
\url{https://aisel.aisnet.org/amcis2020/strategic_uses_it/strategic_uses_it/12}

Lis, D. and Otto, B., (2021), January. Towards a Taxonomy of Ecosystem
Data Governance. In~\emph{Proceedings of the 54th Hawaii International
Conference on System Sciences}~(p. 6067).

Marelli, L., Testa, G., \& Van Hoyweghen, I. (2021). Big Tech platforms
in health research: Re-purposing big data governance in light of the
General Data Protection Regulation's research exemption.~\emph{Big Data
\&
Society},~\emph{8}(1).~\url{https://doi.org/10.1177/20539517211018783}

Mirowski, P. (2018). The future(s) of open science. Social Studies of
Science, 48(2), 171--203. \url{https://doi.org/10.1177/0306312718772086}

My Health Record. Privacy Policy. (2018).
\url{https://www.myhealthrecord.gov.au/about/privacy-policy}. Accessed
10 February 2020.

Nokkala, T., Salmela, H. and Toivonen, J., (2019). Data governance in
digital platforms. Twenty-fifth Americas Conference on Information
Systems, Cancun, 2019. Available at:
\url{https://web.archive.org/web/20220803150237id_/https://aisel.aisnet.org/cgi/viewcontent.cgi?article=1368\&context=amcis2019}

Office of the Australian Information Commissioner. National Repositories
Service --- eHealth record System Operator: Audit report. (2014).
\url{https://www.oaic.gov.au/privacy/privacy-assessments/national-repositories-service-ehealth-record-system-operator-audit-report/}.
Accessed 10 February 2020.

Parisi, L., (2016). Automated thinking and the limits of reason.
\emph{Cultural Studies/Critical Methodologies}, 16(5), 471-481.

Parisi, L. (2019). The alien subject of AI. \emph{Subjectivity},
\emph{12}(1), 27-48.

Pels, P., Boog, I., Henrike Florusbosch, J., Kripe, Z., Minter, T.,
Postma, M., Sleeboom‐Faulkner, M., Simpson, B., Dilger, H., Schönhuth,
M., Poser, A., Castillo, R. C. A., Lederman, R., \& Richards‐Rissetto,
H. (2018). Data management in anthropology,~Social
Anthropology/Anthropologie sociale,~26(3), 391-413. Retrieved Jan 21,
2023,
from~\url{https://www.berghahnjournals.com/view/journals/saas/26/3/soca12526.xml}

Ringel, L., Hiller, P. \& Zietsma, C. (2018). Toward permeable
boundaries of organizations? \emph{Research in the Sociology of
Organizations}, 47, 3-28.

Scassa, T. (2020). Designing Data Governance for Data Sharing: Lessons
from Sidewalk Toronto.~\emph{Technology and Regulation},~\emph{2020},
44-56.

Social Science One. Social Science One. (2018).
\url{https://socialscience.one/}. Accessed 10 February 2020.

Social Science One. Data Security \& Privacy. (2018).
\url{https://socialscience.one/data-security-privacy}. Accessed 10
February 2020.

Social Science Research Council. Social Science Research Council (SSRC)
\textbar{} Brooklyn, NY, USA. (2018). Retrieved from
\url{https://www.ssrc.org/}. Accessed 10 February 2020.

Social Science Research Council. To Secure Knowledge: Social Science
Partnerships for the Common Good. (2018).
\url{https://s3.amazonaws.com/ssrc-static/tsk/SSRC+To+Secure+Knowledge.pdf}.
Accessed 10 February 2020.

Tallon, P. P. (2013). Corporate governance of big data: Perspectives on
value, risk, and cost. \emph{Computer}, 46(6), 32-38.

Tannam, E. (2018). A guide to the new privacy changes at Facebook.
\url{https://www.siliconrepublic.com/enterprise/facebook-data-privacy-settings}.
Accessed 10 February 2020.

University Medical Center Ultrecht. Work packages. (2017).
\url{https://www.bigdata-heart.eu/About/Work-packages}. Accessed 10
February 2020.

van den Broek, T. \& van Veenstra, A.F. (2015). "Modes of Governance in
Inter-organizational Data Collaborations". ECIS 2015 Completed Research
Papers. Paper 188.

van Thiel, G. An interview with Prof. dr. JJM (Hans) van Delden,
Professor of Medical Ethics at the University Medical Center Utrecht,
and WP 7 co-lead BigData@Heart. (2018).
\url{https://www.bigdata-heart.eu/News/ID/77/Interview-with-Prof-dr-JJM-Hans-van-Delden-governance-ethics-and-legal-aspects-in-the-work-of-BigDataHeart}.
Accessed 10 February 2020.

Zuckerberg, M. The Internet needs new rules. Let's start in these four
areas. (2019)
\url{https://www.facebook.com/zuck/posts/10107013839885441}.
Accessed 10 February 2020.

\end{multicols*}

\pagebreak

\hypertarget{appendix-1-comparison-of-data-governance-partnerships}{%
\subsection{Appendix 1: Comparison of Data Governance
Partnerships}\label{appendix-1-comparison-of-data-governance-partnerships}}

\begin{longtabu}[]{@{}XXXXX@{}}
\toprule
\endhead
& \textbf{SS1} & \textbf{MHR} & \textbf{Big Data @ Heart} &
\textbf{Insight} \\
\emph{Geography} & US, global & Australia & European & Ireland \\
\emph{Field} & Social media & Health & Health & Research (biotech;
ICT) \\
\emph{Key Partners} & Private Companies; Non-profit foundations;
Universities & Government departments & Health Corporations (``Big
Pharma''); Universities; Private Research Institutes & Universities,
industry \\
\emph{Funding} & Non-profit foundations & Government (tax revenue) &
Member contributions & University funds; research grants, funding from
industry \\
\emph{Source of data} & Facebook users & Australian citizens (who have
not opted out) & Participants in clinical trials; Health patients &
Varied \\
\emph{User control} & Via Facebook's privacy controls (no \emph{ex post}
consent) & Via MHR's platform (limited) & N/A & N/A \\
\emph{Data access method} & Grants & Via pre-determined roles
(government agencies, hospitals, doctors, nominated family members). MHR
Data Governance Board that evaluates applications for secondary use of
MHR data. & Membership & N/A \\
\emph{Ethical oversight} & University IRBs & For primary use: medical
ethics committees; for secondary use, university ethics committees & N/A
& University IRBs \\
\emph{Governance arbiter or decision-maker} & SS1 commission,
subcommittee & For primary, i.e. clinical use: ADHA; for secondary, i.e.
research use: MHR Data Governance Board + AIHW & N/A & N/A \\
\emph{Governance guidelines or regulation} & King \& Persily's article &
Federal Government Legislation & Work Package 7 & Magna Carta for
Data \\
\emph{Technical model for privacy} & Differential privacy; physical
control over data access; encryption & Encryption & N/A & N/A \\
\emph{Data stored} & Dataverse; Amazon AWS & Data centres operated by
Accenture & N/A & At respective universities if data is collected by
researchers or in the company databases if the research is done on
industry-owned data \\
\emph{Partnership model} & Pre-defined, highly structured and
combinatorial of existing institutions (Harvard as ``incubator'';
nonprofit foundations as funders; other universities, through committee
members) & Defined by legislation; involving all health industry
stakeholders & Consortium; evolving and experimental & Consortium;
evolving and experimental \\
\bottomrule
\end{longtabu}

\end{document}